\documentclass[a4paper]{llncs}
\usepackage[top=2.05in, bottom=2.05in, left=1.73in, right=1.73in]{geometry}
\usepackage{amssymb}
\usepackage{graphicx}

\usepackage{url}
\urldef{\mailsa}\path|{alfred.hofmann, ursula.barth, ingrid.haas, frank.holzwarth,|
\urldef{\mailsb}\path|anna.kramer, leonie.kunz, christine.reiss, nicole.sator,|
\urldef{\mailsc}\path|erika.siebert-cole, peter.strasser, lncs}@springer.com|    
\newcommand{\keywords}[1]{\par\addvspace\baselineskip
\noindent\keywordname\enspace\ignorespaces#1}

\usepackage{tabularx}
\newcolumntype{L}[1]{>{\raggedright\arraybackslash}p{#1}}
\newcolumntype{C}[1]{>{\centering\arraybackslash}p{#1}}
\newcolumntype{R}[1]{>{\raggedleft\arraybackslash}p{#1}}

\usepackage{booktabs}
\usepackage{cite}
\usepackage{subfigure}
\usepackage{amsmath}

\begin{document}

\mainmatter  

\title{Automated retinal vessel segmentation based on morphological preprocessing and 2D-Gabor wavelets}


%
%
\author{Kundan Kumar\inst{1}\thanks{corresponding author} \and Debashisa Samal\inst{1} \and Suraj\inst{2}%
}
%


\institute{Department of Electronics and Communication Engineering, ITER, Siksha `O' Anusandhan (Deemed to be University), Bhubaneswar-751030, Odisha, India\\
\email{erkundanec@gmail.com, debashishsamal@soa.ac.in}
\url{https://sites.google.com/site/erkundanec/home} 
\and Department of Electrical Engineering, Sardar Vallabhbhai National Institute of Technology, Surat, Gujarat-395007\\\email{suraj.boom@gmail.com}}

%
%

\tocauthor{Authors' Instructions}
\maketitle

\begin{abstract}
Automated segmentation of vascular map in retinal images endeavors a potential benefit in diagnostic procedure of different ocular diseases. In this paper, we suggest a new unsupervised retinal blood vessel segmentation approach using top-hat transformation, contrast-limited adaptive histogram equalization (CLAHE), and 2-D Gabor wavelet filters. Initially, retinal image is preprocessed using top-hat morphological transformation followed by CLAHE to enhance only the blood vessel pixels in the presence of exudates, optic disc, and fovea. Then, multiscale 2-D Gabor wavelet filters are applied on preprocessed image for better representation of thick and thin blood vessels located at different orientations. The efficacy of the presented algorithm is assessed on publicly available DRIVE database with manually labeled images. On DRIVE database, we achieve an average accuracy of 94.32\% with a small standard deviation of 0.004. In comparison with major algorithms, our algorithm produces better performance concerning the accuracy, sensitivity, and kappa agreement.
\end{abstract}

\section{Introduction}
Change in anatomical structure of retinal blood vessels (vasculature) in retina is a good indication of the presence of ophthalmic diseases, e.g., hypertension, cardiovascular diseases, diabetic retinopathy,  glaucoma, etc.~\cite{Soares2006, Zhao2014}. Vascular map segmentation of fundus images has played a decisive role in assessing the change in the vasculature for severity of ocular diseases. However, periodic screening of retinal images for early recognition of the change in vascular structure can prevent major vision loss~\cite{Zhao2014}. Therefore, an automatic and accurate retinal vessel segmentation is the prerequisite for the initial diagnosis of retinal diseases. Extraction of the vascular map from an uneven illuminated and pigmented fundus image is a challenging problem. Besides retinal blood vessels, the presence of other structures (e.g., exudates, optic disc, fovea, red lesions) under uneven illuminated and pigmented background makes the vessel detection even more difficult. Also, the retinal vessel thickness varies in the wide range whereas thin vessels have low contrast which makes thin vessel detection more challenging~\cite{Zhao2014}.

\par Several, automatic vascular map segmentation of retinal fundus images have been proposed in literature~\cite{Chaudhuri1989, Hoover2000, Zhang2010, Liskowski2016, Gou2018}. Fraz \textit{et al.} have done a comprehensive literature survey on retinal blood vessel segmentation in~\cite{Fraz2012a}. The retinal blood vessel segmentation techniques are popularly classified as: (i) supervised and (ii) unsupervised techniques. In supervised category, $k$-NN-classifier~\cite{Staal2004}, Artificial Neural Network (ANN)~\cite{Franklin2014}, trainable COSFIER filters~\cite{Azzopardi2015}, Support Vector Machine (SVM)~\cite{Ricci2007}, Extreme Learning Machine (ELM)~\cite{Zhu2017}, Deep Neural Network (DNN)~\cite{Liskowski2016, Sadek2017}, etc. have been explored for blood vessel segmentation as identification problem. However, supervised algorithms rely on the robust feature extraction followed by classification. In many approaches line detector~\cite{Ricci2007}, Gabor filters~\cite{Soares2006,Franklin2014}, and gray level co-occurrence matrix (GLCM)~\cite{Rahebi2014} based methods have been explored for feature extraction purpose. A feature vector is computed for each pixel using these approaches and classifier classify the pixels as the vessel and non-vessel pixels. These approaches are time-consuming process due to training. On the contrary, in the unsupervised category, filtering methods~\cite{Chaudhuri1989, Zhang2010, Li2012}, vasculature tracing methods~\cite{Sofka2006}, curvelet based~\cite{Miri2011}, morphological operators \cite{Hassan2015}, have been used. In these approaches, classification of the vessel or non-vessel pixels is performed without training process, i.e., training data do not contribute to finding the model parameter.

\par In earlier reported works under the unsupervised category, match filter has received the enormous response of the scholars due to its straightforwardness in the implementation of the technique. Match filter based retinal vessel segmentation relies on a 2D kernel with Gaussian profile initially proposed by Chaudhuri \textit{et al.}~\cite{Chaudhuri1989}. The kernel is rotated at $15^o$ increment, and the best output of the filter for each pixel is carefully chosen to map all blood vessels orientated at different angles. After that, thresholding is applied to get binary vessel map image. Further, pruning is applied as post-processing to improve the final identification of blood vessels. Hoover \textit{et al.}~\cite{Hoover2000} have used local and region-based properties for vessel segmentation where threshold probing technique is used on match filter response. However, match filter gives a strong response in terms of vessels and non-vessels edges. Zhang \textit{et al.}~\cite{Zhang2010} have exploited the first-order derivative of Gaussian to improve the performance of matched filter by eliminating non-vessel edges from retinal images. In literature~\cite{Sofka2006}\cite{Li2012}, multiscale match filter and its variation are suggested to identify blood vessels of different thickness. In~\cite{Zhao2014}, Zhao  \textit{et al.} have enhanced the retinal vessels in retinal images by utilizing 2D-Gabor wavelet filters and a contrast-limited adaptive histogram equalization (CLAHE). After that, the processing results of region growing method and level set approach are combined to get final segmentation as a binary image. However, this approach is unable to remove non-vessel structures also takes long processing time. Roychowdhury \textit{et al.}~\cite{Roychowdhury2015} have performed an unsupervised iterative process to obtain the vessels using top-hat reconstruction followed by iterative region growing method. Most of the approaches like~\cite{Zhang2010, Zhao2014} fail to remove optic disc and exudates from pathological images. Also, many approaches remove these structures in postprocessing with the extra burden of computational time. Thus, an automated unsupervised blood vessel segmentation approach is needed to identify the blood vessel pixels correctly with a small false positive rate. Simultaneously, need to remove the anatomical structure other than blood vessels with high accuracy and less complexity.

\par We propose an entirely unsupervised approach for automatic segmentation of blood vessels to obtain the retinal vascular map. The significant contribution of this paper is to use top-hat transform followed by CLAHE for retinal image enhancement in preprocessing step. Use of top-hat transform facilitates to enhance only the blood vessels, simultaneously remove the local intensity change due to exudates, optic disc, and fovea from the background. For further image enhancement, CLAHE is applied on top-hat transformed retinal image. The preprocessed retinal image is passed through the Gabor filters bank, and the maximum outcome of the filters are chosen for each pixel. Otsu thresholding as global thresholding is applied to get a binary blood vessel structure. The proposed technique is proficient in identifying the blood vessels under uneven pigmentation and illuminance condition in the presence of exudates, optic disc, and fovea. Our proposed approach suppresses the non-vessel pixels in the preprocessing step, however multiscale Gabor wavelet filters efficiently represent the thick and thin vessels in the retinal image. The efficacy of the presented technique is validated on the publicly available Digital Retinal Image for Vessels Extraction (DRIVE) database~\cite{Staal2004}.
\par Rest of the paper is organized as follows. The DRIVE database and the proposed algorithm are discussed in sections~\ref{mater} and \ref{Prop} respectively. The section~\ref{results} discusses the experimental results and finally, the paper is concluded in section~\ref{conclusion}.
\section{Materials}
\label{mater}
We validate our proposed algorithm on the DRIVE database \cite{Staal2004} for performance evaluation. The DRIVE database is publicly available to execute a comparative study and experimental evaluation of vascular segmentation algorithms. The gold standard segmented images are provided with the database as manually labeled images. The DRIVE database contains 40 color retinal images which include 20 images in training set and 20 images in the test set. All images were captured by Canon CR5 nonmydriatic 3 charge-coupled-device (CCD) cameras at $45^o$ field of view (FOV). Each color retinal image is having a resolution of $565\times 584$ pixels with three R, G, and B channels, and each channel is an 8-bit grayscale image. In this paper, we examined our presented algorithm on images from the test set. For test set, two subsets of manually segmented images, i.e., set A and set B, are provided. Set A as the first observer's manual segmented images are considered as gold standard which are utilized as ground truth for performance assessment. The primary objective of choosing DRIVE database is to perform a comparative analysis of the presented work with the state-of-art techniques which have been evaluated on the same database.

\section{Proposed method}
\label{Prop}
\subsection{overview}
\label{overview}
\begin{figure}[!h]
\vspace{-0.5cm}
\centering
\includegraphics[scale=0.16]{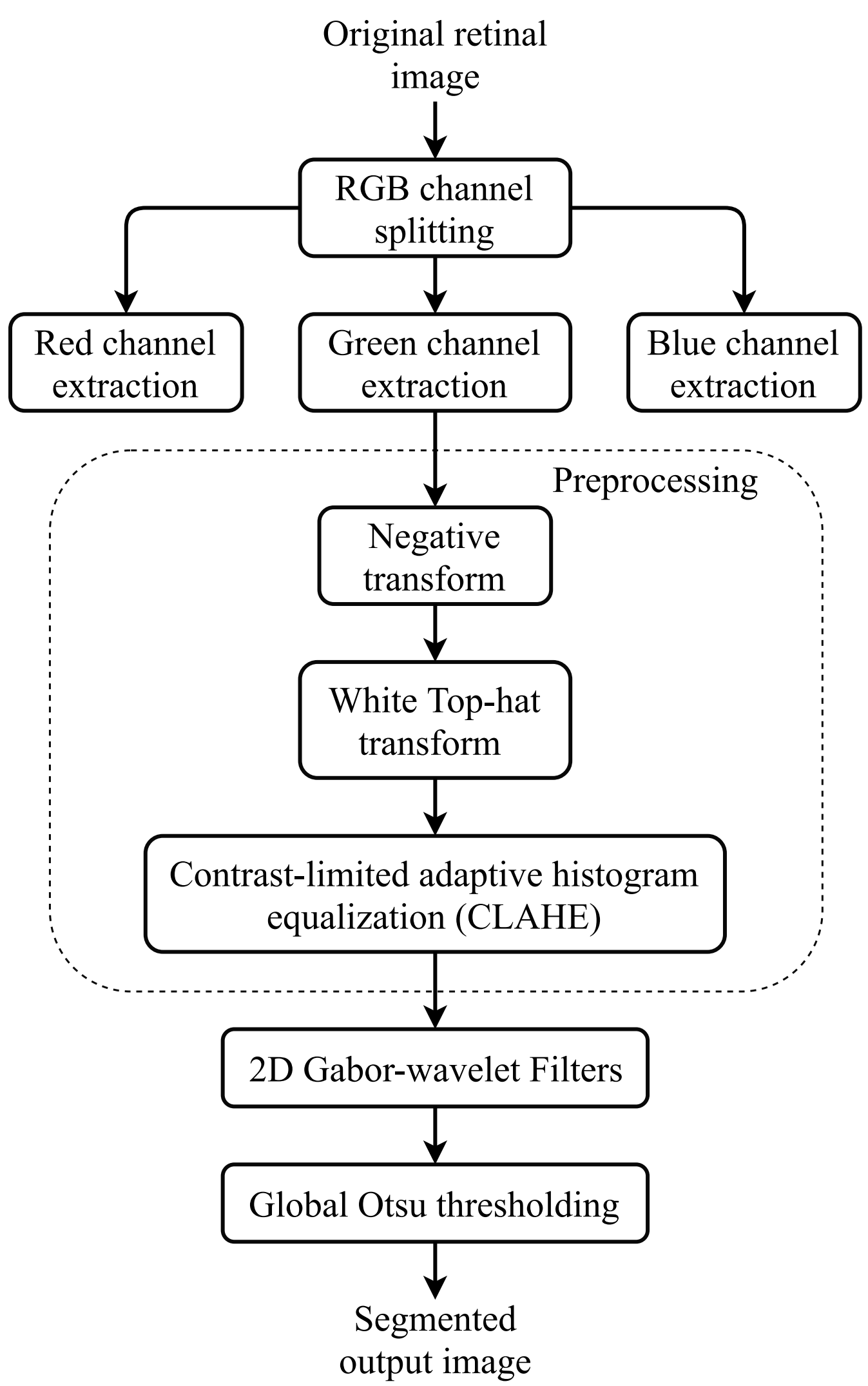}
\vspace{-0.3cm}
\caption{Flow diagram of the proposed algorithm}
\label{blockdiagram}
\vspace{-0.5cm}
\end{figure}

In the presented work, an unsupervised blood vessel segmentation approach is proposed. Fig.~\ref{blockdiagram} presents the flow diagram of the proposed algorithm. 

\par The principal idea of the proposed method is that in retinal image, vessels can be distinguished from other structures like exudates, optic disc, fovea, etc. during preprocessing. In many approaches, CLAHE is employed in preprocessing to boost the dynamic range of the retinal images besides preventing the over-amplification of noise~\cite{Zhao2014}. However, CLAHE improves the contrast of images by operating on local regions rather than globally due to which vessel pixels enhance with the other structures too. Therefore, we first applied the top-hat transformation on the green channel of color retinal fundus image that intensifies only the blood vessel pixels and simultaneously suppress the other structure pixels. After that, CLAHE is applied to get the full advantage of its characteristics. In the second stage, 2D-Gabor wavelet filter is applied to the preprocessed image. Furthermore, a global Otsu thresholding method is used to get a binary segmented image.

\subsection{Preprocessing}
\label{preprocessing}
\par
\begin{figure}[!h]
\vspace{-0.5cm}
\centering
\subfigure[]{\includegraphics[height = 2.3cm]{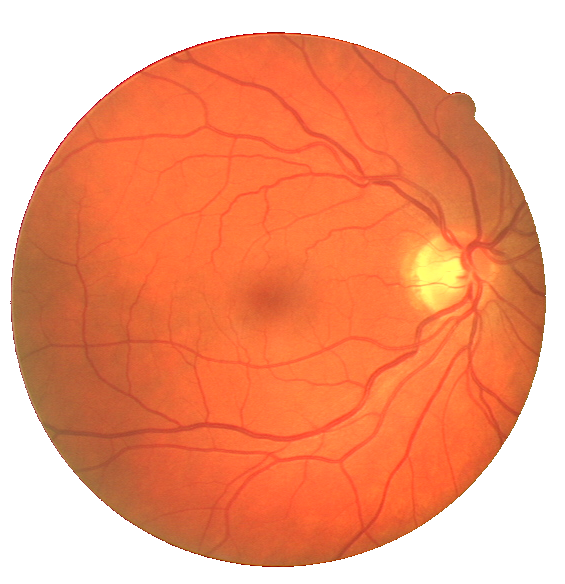}\label{original}}~~~~
\subfigure[]{\includegraphics[height = 2.3cm]{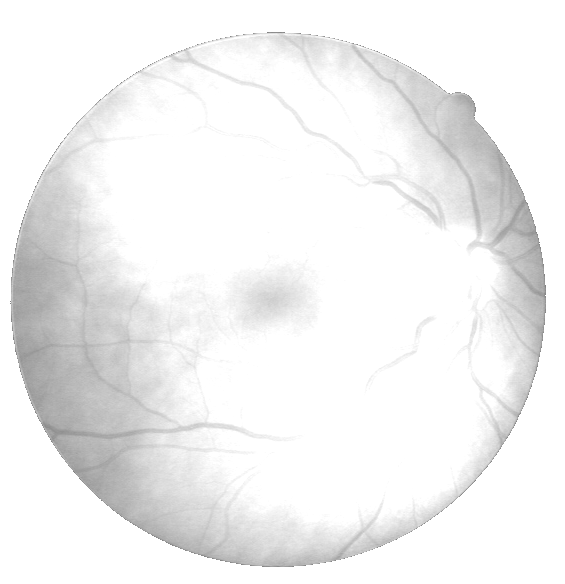}\label{redChannel}}~~~~
\subfigure[]{\includegraphics[height = 2.3cm]{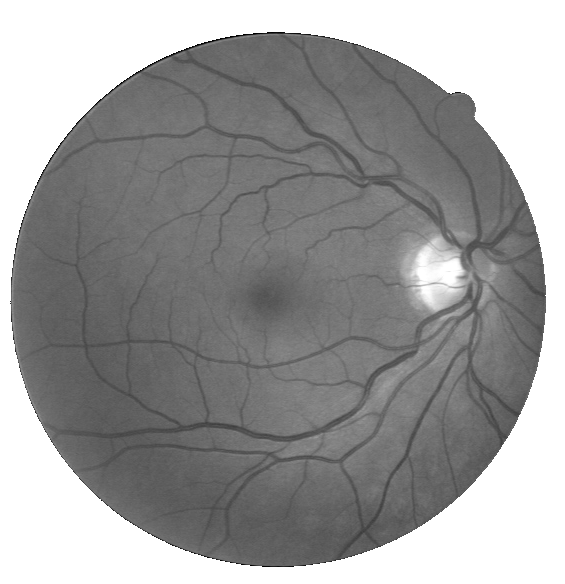}\label{greenChannel}}~~~~
\subfigure[]{\includegraphics[height = 2.3cm]{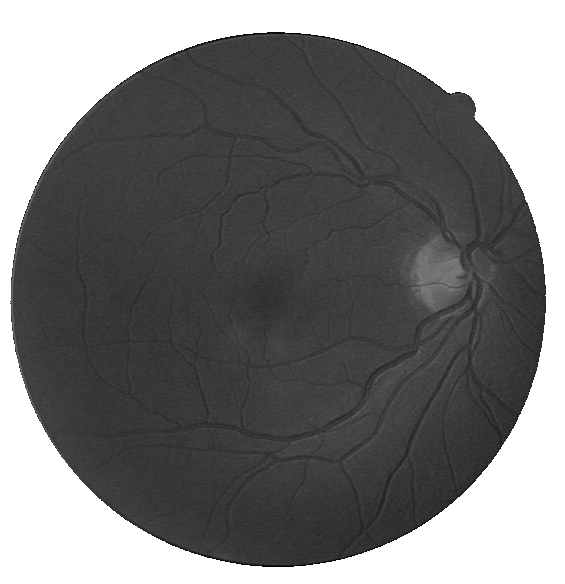}\label{blueChannel}}
\vspace{-0.4cm}
\caption{Retinal image through FOV: (a) original RGB color image, (b) Red channel, (c) green channel, and (d) blue channel.}
\label{split}
\vspace{-0.5cm}
\end{figure}

 Initially, an original RGB color retinal image is split into three channels as shown in Fig.~\ref{split}.
Among these three channels, the green channel image appears having higher contrast compared to other two channel images. In green channel image, the blood vessel pixels are visible and easily distinguishable from the background pixels. Because in our eyes, lens pigments absorb light colors differently~\cite{Walter2007, Zhao2014}. Therefore, red vessels in color retinal image are more visible in the green channel image as presented in Fig.~\ref{greenChannel}, however, red channel image is the brightest image and blue channel image suffers from poor dynamic range as shown in Fig.~\ref{split}.

\begin{figure}[!h]
\vspace{-0.5cm}
\centering
\subfigure[]{\includegraphics[height=2.3cm]{img_02_test.png}}~~~~~~~
\subfigure[]{\includegraphics[height=2.3cm]{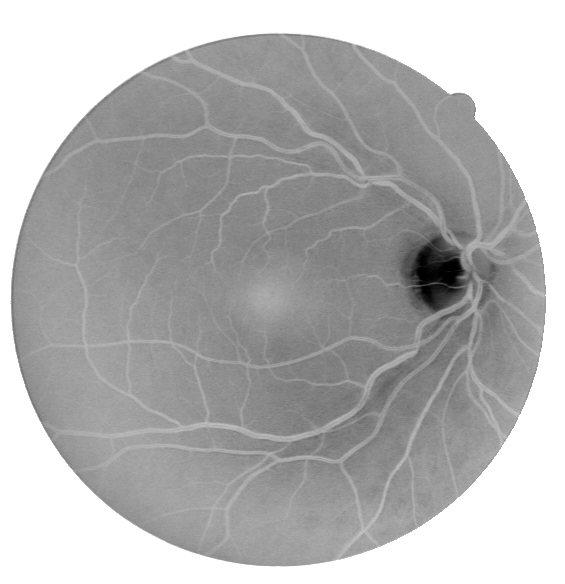}\label{InvGreen}}~~~~~~~
\subfigure[]{\includegraphics[height=2.3cm]{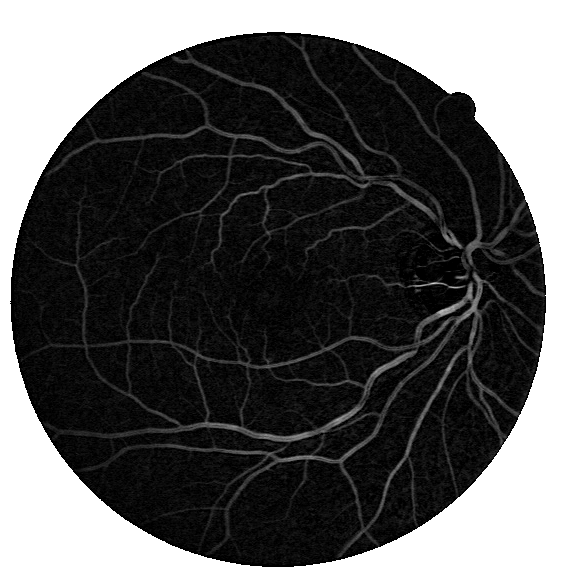}\label{imgGTopHat}}\\
\subfigure[]{\includegraphics[height=2.3cm]{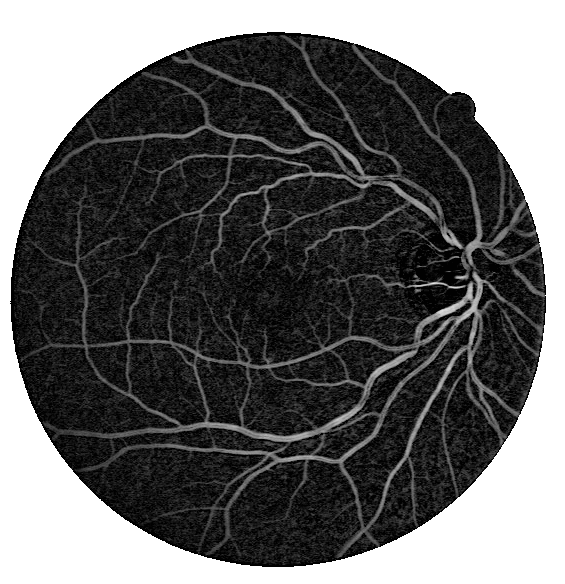}\label{imgGClahe}}~~~~~~~
\subfigure[]{\includegraphics[height=2.3cm]{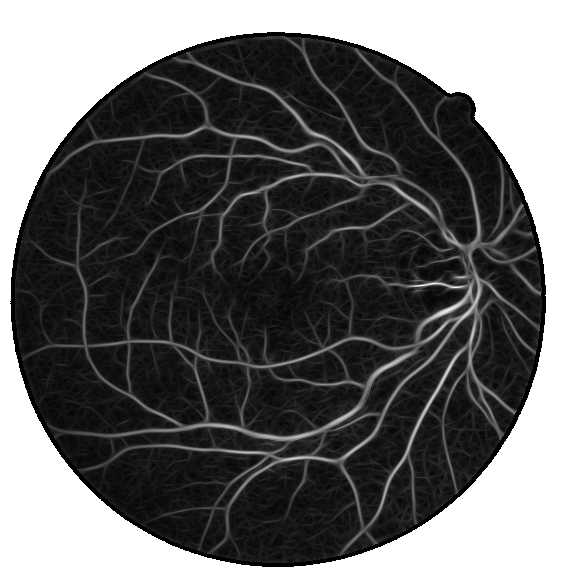}\label{imgGaborR}}~~~~~~~
\subfigure[]{\includegraphics[height=2.3cm]{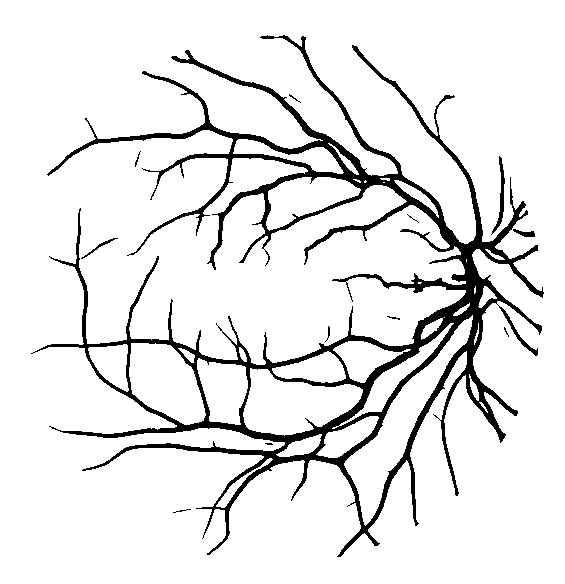}\label{binary}}
\vspace{-0.4cm}
\caption{Processing results at each intermediate steps of the proposed algorithm. (a) Original test image, (b) Inverted green channel image, (c) white top-hat transformed image of inverted green channel image, (d) CLAHE processed image, (e) Gabor wavelet response, (f) Binary image after global thresholding.}
\label{completStep}
\vspace{-0.5cm}
\end{figure}
In the green channel image ($I_G$), the blood vessel pixels due to its intensities being close to 0 seems to be dark.  Image $I_G$ is inverted followed by superposition of fundus mask ($M$) to make the blood vessel pixels brighter and keep the focus on the region of interest (FOV). The inverted green channel image through fundus mask is shown in Fig.~\ref{InvGreen}. To enhance only the vessels, white top-hat transformation is applied on inverted green channel image ($I'_G$). Usually, blood vessels have small thickness compared to the other structures in retinal images. Therefore, a circular structuring element of diameter at least equal to the diameter of the thickest blood vessel is preferred for top-hat transformation. The white top-hat transform can be defined as
\begin{equation}
T_w(I'_G) = I'_G-I'_G \circ b.
\end{equation}

Where $T_w$ is the transformed image of $I'_G$ using structuring element $b$, and $\circ$ denote the morphological opening operator. In image processing, the top-hat transformation is a morphological operation which highlights the object smaller than the structuring element~\cite{Dougherty2003}. The diameter of the structuring element is chosen 11 pixels wide as the maximum width of the blood vessel is less than 11 pixels. For the DRIVE database, the width of the blood vessel varies in the range of 1-10 pixels~\cite{Fathi2013}. The diameter of the structuring element may vary for the different database having different image resolution.

Usually, the width of widest blood vessels is less than the width of other structures, like exudates, fovea, and optic disc, which do not appear in the top-hat transformed image ($T_w$) as shown in Fig.~\ref{imgGTopHat}. Also, a homogeneous background is obtained in the transformed image. After applying the top-hat transformation, the blood vessel pixels do not achieve a good contrast compared to the background. Therefore, (CLAHE) technique is adopted for further enhancement of the processed retinal image. Fig~\ref{imgGClahe} shows the CLAHE response ($I_c$) of the top-hat transformed image (Fig.~\ref{imgGTopHat}). However, CLAHE also enhances the background noise. For informative representation of the blood vessels having different thickness and orientation, retinal image is processed through the multiscale Gabor wavelet filters at different frequencies and orientations. Gabor wavelet also smoothes the background noise.

\subsection{2D-Gabor wavelet filter bank}
\label{gabor}
The 2D-Gabor wavelet transformation is a tool for a complete representation of an image in terms of radial frequency and orientation~\cite{Lee1996}. To highlight the blood vessels of different width placed at different orientations in this work, we used a bank of 2D-Gabor wavelet filters to the preprocessed retinal image. Daugman \textit{et al.}~\cite{Daugman1988} have proposed that ensemble of a simple cell of visual cortex can be represented as a family of 2D-Gabor wavelets. The decomposition of an image, $f=I_c$, as wavelet transform is defined as
\begin{equation}
({T^{wav}}f)(a,\theta ,{x_0},{y_0}) = {\left\| a \right\|^{ - 1}}\iint {dxdyf(x,y){\psi _\theta }\left( {\frac{{x - {x_0}}}{a},\frac{{y - {y_0}}}{a}} \right)},
\label{waveletTrans}
\end{equation}
where $\theta$ is the orientation parameter of the wavelet, and $a$ is the parameter that defines the standard deviation in $x$ and $y$ directions. In equation~(\ref{waveletTrans}),
\begin{equation}
{\psi _\theta }(a,x,y,{x_0},{y_0}) = {\left\| a \right\|^{ - 1}}{\psi _\theta }\left( {\frac{{x - {x_0}}}{a},\frac{{y - {y_0}}}{a}} \right)
\label{motherWavelets}
\end{equation}
represents the elementary function of the 2D wavelet rotated by an angle $\theta$. Using the Gabor elementary function, the entire family of Gabor wavelets can be generated. Lee \textit{et al.}~\cite{Lee1996} have derived a specific class of 2D-Gabor wavelets which is used in this paper to obtain a set of Gabor wavelets to process the retinal images. The Gabor wavelet which satisfies the neurophysiological restraint of simple cells is defined as
\begin{footnotesize}
\begin{align}
\psi (x,y,{\omega _0},\theta,K ) = &\frac{{{\omega _0}}}{{\sqrt {2\pi } K }}\exp \left( { - \frac{{\omega _0^2}}{{8{K^2}}}\left( {4{{(xcos\theta  + y\sin \theta )}^2} + {{( - xsin\theta  + ycos\theta )}^2}} \right)} \right)\\
&\cdot \left[ {\exp \left\{ {i{\omega _0}\left( {xcos\theta  + y\sin \theta } \right)} \right\} - \exp \left( { - \tfrac{{{K^2}}}{2}} \right)} \right]\nonumber
\end{align}
\end{footnotesize}
where, $\theta$ denotes the wavelet orientation in radians and $\omega_0$ is the radial frequency in radian per unit length. The constant $K$ tells about the frequency bandwidth of octave where $K=\pi$ is for a frequency bandwidth of one octave and $K\approx 2.5$ for frequency bandwidth of 1.5 octaves. Each Gabor wavelet filter at radial frequency $\omega_0$ and orientation $\theta$ is centered at $(x=0,y=0)$ and normalized by $L^2$ norm.
\par
\begin{figure}[!h]
\vspace{-0.5cm}
\centering
\subfigure[$\omega_0=0.7$]{\includegraphics[scale=1.6]{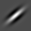}}~~~~~~~~
\subfigure[$\omega_0=0.9$]{\includegraphics[scale=1.6]{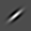}}~~~~~~~~
\subfigure[$\omega_0=1.1$]{\includegraphics[scale=1.6]{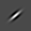}}~~~~~~~~
\subfigure[$\omega_0=1.3$]{\includegraphics[scale=1.6]{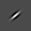}}
\vspace{-0.3cm}
\caption{Gabor wavelet filters at $50^o$ orientation for four different radial frequencies}
\label{gaborFilters}
\vspace{-0.6cm}
\end{figure}

 For each pixel in the retinal image, maximum Gabor wavelet outcome over all possible filters is stored for filtered image. If we consider $T_\psi (\omega_0,\theta,K)$ as the transformed retinal image at angular frequency $\omega_0$ and orientation $\theta$. Then Gabor wavelet transformation result is obtained as 
\begin{equation}
P(K) = \mathop {max}\limits_{{\omega _0},\theta } \left| {{T_\psi }({\omega _0},\theta ,K)} \right|
\end{equation}
where $\theta$ is varied in the range of $[0,180)$ at an equal interval of $20^\circ$. The radial frequency is varied between $[0.7,1.5)$ at an interval of $0.2$. Lee \textit{at el.}\cite{Lee1996} have suggest to choose $K$ in the range of [2,2.5]. In this paper, $K=2.2$ is selected to accomplish better distinguishability between background and vessels. All these parameters are selected by performing few experiments on retinal images and Gabor wavelet filters. Fig.~\ref{gaborFilters} shows few Gabor wavelet filters from the bank of filters for different radial frequency at orientation of $50^o$. The Gabor wavelet filter at $\omega_0=0.7$ is efficient to detect thick blood vessel. However, $\omega_0=1.3$ is suitable to detect thin blood vessels. The Gabor wavelet response of the preprocessed image is shown in Fig.~\ref{imgGaborR}.

The hard segmented binary image as illustrated in Fig.~\ref{binary} is obtained by applying Otsu thresholding on the filtered image $P$.
\section{Results and Discussions}
\label{results}
The presented algorithm is evaluated and compared with competitive algorithms using five different metrics: accuracy ($Acc$), sensitivity ($Se$), specificity ($Sp$),  kappa agreement ($\kappa$), and area under the curve ($A_z$). All the metrics are computed using only pixels inside the FOV. Accuracy, sensitivity, and specificity are calculated using false positive ($X$), false negative ($Y$), true positive ($Z$), and true negative ($W$) values. $Z$ denotes the number of pixels correctly identified as vessel pixels, and $X$ denotes the number of pixels belongs to the background but wrongly identified as vessel pixel. $W$ represents the number of pixels correctly identified as background pixels, and $Y$ represents the number of pixels belongs to the vessel but incorrectly assigned to background pixels. The evaluation metrics are measured using the following mathematical expressions as
\begin{footnotesize}
\begin{equation}
Acc =\frac{Z+W}{Z+X+Y+W}, ~~~~~~ Se = \frac{Z}{Z+Y},
\end{equation}
\begin{equation}
Sp = \frac{W}{X+W},   ~~~~~~~~~~~ \kappa = \frac{p_o-p_e}{1-p_e}.
\end{equation}
\end{footnotesize}

Where $p_e$ is the hypothetical probability of chance agreement and $p_o$ is the relative perceived agreement. These agreement values can be calculated using the perceived data to estimate the probabilities of each observer randomly seeing all the classes. $\kappa$ value varies in the range of $[0,1]$, where $\kappa=0$ relates to no agreement between two rates, whereas $\kappa=1$ relates to complete agreement between the rates. To compute $A_z$, receiver operating characteristic (ROC) curve is attained by changing the global threshold between 1 and 0 in steps of 0.01. The global threshold divides the image into a binary image with labels 0 and 1. For each threshold, two performance measure false positive rate ($XR =  1-Sp$) and true positive rate ($ZR = Sp$) are obtained by comparing the segmented binary image with the corresponding ground truth. Before applying the global threshold, the Gabor wavelet response values are normalized to 0-1 range.

\begin{figure}[!h]
\vspace{-0.5cm}
\centering
\subfigure[]{\includegraphics[height=6cm]{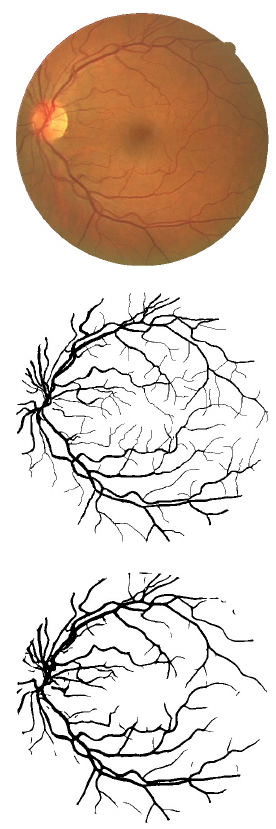}\label{seg01}}~~~~
\subfigure[]{\includegraphics[height=6cm]{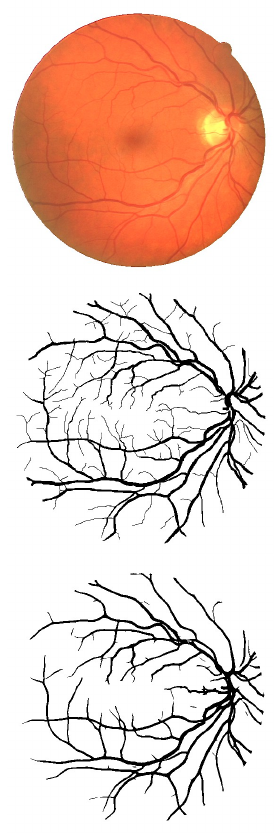}\label{seg02}}~~~~
\subfigure[]{\includegraphics[height=6cm]{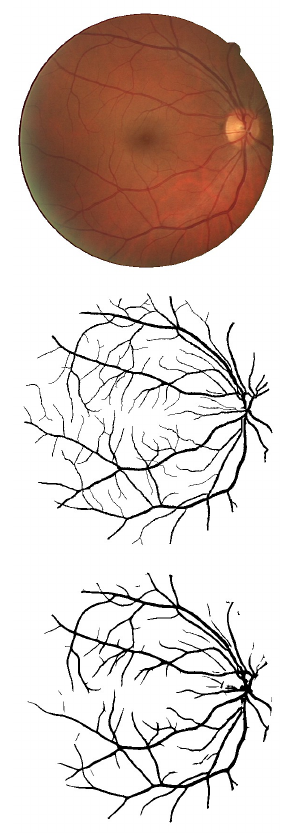}\label{seg03}}~~~~
\subfigure[]{\includegraphics[height=6cm]{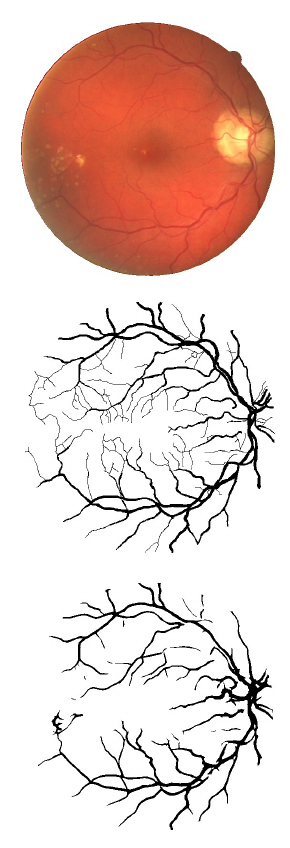}\label{seg04}}
\vspace{-0.4cm}
\caption{Segmentation result of retinal images from the DRIVE database. First row: original image, Second row: golden standard ground truth, Third row: segmented vessel using the proposed algorithm. (a-c) Healthy retinal image with different pigmentation and illuminance, (d) Pathological retinal image.}
\label{finalResult}
\vspace{-0.5cm}
\end{figure}

The blood vessel segmentation outcome performances are shown in Fig.~\ref{completStep}. All intermediate results and segmented output of the presented algorithm for a retinal image from DRIVE database are illustrated in Fig.~\ref{completStep}. The segmentation performance of the presented algorithm on the DRIVE test data set is listed in Table~\ref{compResults}.
\renewcommand{\arraystretch}{0.8}
\begin{table}[!h]
\vspace{-0.5cm}
\caption{Segmented outcome of the presented work on the DRIVE test database}
\begin{center}
\begin{scriptsize}
\vspace{-0.2cm}
\begin{tabular}{C{1cm}C{1.7cm}C{1.7cm}C{1.7cm}C{1.7cm}C{1.7cm}}
\toprule
\multicolumn{1}{l}{Image} & Se & Sp & Acc & $A_z$ & $\kappa$\\ \midrule
1 & 0.8337  & 0.9592 & 0.9427 & 0.9687 & 0.7593 \\
2 & 0.7781  & 0.9767 & 0.9469 & 0.9602 & 0.7837 \\
3 & 0.7583  & 0.9639 & 0.9339 & 0.9432 & 0.7312 \\
4 & 0.6619  & 0.9874 & 0.9440 & 0.9443 & 0.7282 \\ 
5 & 0.7010  & 0.9837 & 0.9453 & 0.9468 & 0.7460 \\ 
6 & 0.6975  & 0.9784 & 0.9388 & 0.9371 & 0.7279 \\ 
7 & 0.7068  & 0.9737 & 0.9383 & 0.9431 & 0.7171 \\
8 & 0.7111  & 0.9721 & 0.9393 & 0.9395 & 0.7120 \\
9 & 0.7289  & 0.9771 & 0.9480 & 0.9525 & 0.7378 \\ 
10 & 0.7021 & 0.9816 & 0.9483 & 0.9507 & 0.7354 \\ 
11 & 0.7378 & 0.9660 & 0.9365 & 0.9412 & 0.7142 \\ 
12 & 0.7936 & 0.9618 & 0.9408 & 0.9517 & 0.7364 \\ 
13 & 0.6907 & 0.9798 & 0.9388 & 0.9481 & 0.7273 \\ 
14 & 0.8094 & 0.9612 & 0.9433 & 0.9581 & 0.7386 \\ 
15 & 0.7800 & 0.9672 & 0.9477 & 0.9566 & 0.7268 \\
16 & 0.7225 & 0.9787 & 0.9451 & 0.9632 & 0.7440 \\ 
17 & 0.7440 & 0.9684 & 0.9407 & 0.9451 & 0.7221 \\ 
18 & 0.7920 & 0.9629 & 0.9432 & 0.9613 & 0.7300 \\
19 & 0.8476 & 0.9713 & 0.9564 & 0.9729 & 0.7991 \\ 
20 & 0.8080 & 0.9623 & 0.9459 & 0.9638 & 0.7306 \\\midrule
\multicolumn{1}{l}{Average} & 0.7503 & 0.9717 & 0.9432 & 0.9524 & 0.7374 \\
\multicolumn{1}{l}{Std. deviation} & 0.0499 & 0.0081 & 0.0049 & 0.0098 & 0.0206 \\ \midrule
\multicolumn{6}{L{11cm}}{Se - sensitivity, Sp - specificity, Acc - accuracy, $A_z$ - Area under the ROC curve, $\kappa$ - Kappa agreement}\\\bottomrule
\end{tabular}
\vspace{-1cm}
\end{scriptsize}
\end{center}
\label{compResults}
\end{table}
The performance metrics illustrate that average segmentation accuracy, sensitivity, and specificity of the presented technique are 94.32\%, 75.03\%, and 97.12\% respectively. It can be noticed that a small deviation in accuracy ($\sigma = 0.0049$) is achieved. In DRIVE test data set, image 8 is a pathological image having exudates on which we have achieved 93.93 \% accuracy with 0.7111 $ZR$. However, Zhao et al.~\cite{Zhao2014} have reported 93.59 \% of accuracy with 0.6524 $ZR$ on the same image. The segmented blood vessels of image 8 is presented in Fig.~\ref{seg04}. It confirms that our proposed algorithm is efficient in segmenting the blood vessels in pathological images also. Average area under the curve, $A_z$, of 0.9524 is obtained with maximum and minimum $A_z$ as 0.9729 and 0.9371 respectively. On DRIVE test data set, we achieved average kappa agreement 0.7374 with 0.0206 standard deviations.

\par A comparative analysis of the presented algorithm with competing methods is presented in Table~\ref{compMethod} in terms of average accuracy, sensitivity, specificity, $A_z$, and $\kappa$ agreement. The evaluation metrics values of different methods provided in the table are taken from the respective papers. Numerical results show that our proposed algorithm outperforms many unsupervised methods to DRIVE dataset concerning accuracy and sensitivity. Furthermore, the proposed algorithm is better than few supervised techniques on DRIVE test dataset concerning sensitivity and specificity.
\begin{table}[!h]
\vspace{-0.4cm}
\caption{Comparative analysis of the presented work with existing approach on the DRIVE test data set with respect to golden standard ground truth image}
\begin{center}
\begin{scriptsize}
\vspace{-0.2cm}
\begin{tabular}{L{3.6cm}C{1.3cm}C{1.3cm}C{2.6cm}C{1.3cm}C{1.3cm}}
\toprule
\multicolumn{1}{l}{Methods}     & Se    & Sp   & Acc  ($\sigma$)           &   $A_z$   & $\kappa$ \\ \midrule\midrule
\multicolumn{6}{L{11cm}}{\textbf{Supervised Methods}}\\\midrule
Soares \textit{et al.}~\cite{Soares2006} & 0.7230 & 0.9762 & 0.9466~~~~~~~~(-) &  0.9614  & -       \\
Staal \textit{et al.}~\cite{Staal2004}   & 0.7194 & 0.9773 & 0.9441~~~~~~~~(-) &  0.9520  & -       \\
Ricci \textit{et al.}~\cite{Ricci2007}   &   -    &      - & 0.9595~~~~~~~~(-) &  0.9558  & -       \\
Lahiri \textit{et al.}~\cite{Lahiri2016a}&  -     &    -   & 0.9530 (0.0030)   &      -   & 0.7090  \\ 
\midrule\midrule
\multicolumn{6}{L{11cm}}{\textbf{Unsupervised Methods}}\\\midrule
2nd observer                    & 0.7760 & 0.9725 & 0.9473~~~~~~~~(-) &    -     & 0.6970   \\
Chaudhuri \textit{et al.}\cite{Chaudhuri1989}$^{*}$ &    -    &    -    & 0.8773 (0.0232)  &  0.7878   & 0.3357     \\ 
Zana \textit{et al.}~\cite{Zana2001}$^{*}$   &    -   &     -   & 0.9377 (0.0077) & 0.8984   & 0.6971   \\ 
Martinez-Parez \textit{et al.}      \cite{Martinez-Perez2007} &     0.7246    &   0.9655     & 0.9344~~~~~~~~(-) & -        & -   \\
Zhang \textit{et al.}~\cite{Zhang2010} &   0.7120   &   0.9724   & 0.9382~~~~~~~~(-)          &    -     & - \\
Miri et.al.~\cite{Miri2011}     & 0.7352 & 0.9795 & 0.9458~~~~~~~~(-)           & -        & -    \\
Zhao \textit{et al.}~\cite{Zhao2014} &       0.7354 &    0.9789    & 0.9477~~~~~~~~(-)           & -        & -    \\ Gou \textit{et al.}
\cite{Gou2018} &       0.7526 &    0.9669    & 0.9393~~~~~~~~(-)           & -        & -    \\ 
\textbf{Presented method}                 & 0.7503 & 0.9717 & 0.9432 (0.0049)          & 0.9524 & 0.7374   
\\ 
\midrule
\multicolumn{6}{L{11cm}}{$^{*}$results are taken from ~\cite{Niemeijer2004}}\\\bottomrule
\end{tabular}
\vspace{-1cm}
\end{scriptsize}
\end{center}
\label{compMethod}
\end{table}
Fig.~\ref{finalResult} shows the segmentation result of four retinal vessel images selected from the DRIVE test data sets. Among these four retinal images, three retinal images (Fig.~\ref{finalResult}(a)-(c)) are healthy retinal images with optic disc and fovea. Whereas, Fig.~\ref{seg04} is the pathological image having exudates, optic disc, and fovea. It can be observed that our proposed algorithm is competent in identifying thin as well as thick vessels in the presence of exudates, fovea, and optic disc. Besides, the proposed algorithm also preserves the connectivity of the blood vessels. The segmentation performance can be further improved by using some local adaptive thresholding technique instead of global thresholding to detect 1-2 pixel thick vessels.
\section{Conclusions}
\label{conclusion}
Blood vessel segmentation in fundus images plays a vital role to detect different ocular diseases. To distinguish blood vessel from other structure like exudates, optic disc, fovea, etc., under uneven illuminance condition is difficult. In this paper, we have been proposed an entirely unsupervised approach for blood vessel segmentation using the retinal image and validated on DRIVE database. In preprocessing, retinal image is enhanced in the two-steps. In the first step, white top-hat transform is applied to enhance only the blood vessels and concurrently suppress all other structures which are larger than structuring element. In next step, the processed image is further enhanced using CLAHE algorithm. After preprocessing, multiscale Gabor wavelet filters are applied to emphasize the thick and thin vessels in the retinal image. A global threshold is obtained from the filtered image by using Otsu's thresholding technique. Experimental results clearly indicate that the proposed technique is efficient to recognize the blood vessels in the presence of exudates, fovea, and optic disc with the average accuracy of 94.32\% with a small standard deviation of 0.0049. Moreover, the suggested algorithm in this paper is simple and easy to implement.

\par The outcome result of the proposed technique depends on the diameter selected as structuring element for the top-hat transform. If some other structures (like small microaneurysms) which look like vessels and have a thickness less than the width of the structuring element then our proposed method may fail to remove those structures in the segmented image. In future, the shape feature can be incorporated with Gabor wavelet response to discarding such kind of the small structures. Furthermore, testing is to be performed on different retinal databases like STARE and CHASE{\_}DB1 to test the robustness of the proposed method.

\bibliographystyle{splncs03_unsrt}
\bibliography{mybib.bib}
\end{document}